\documentclass{article}

\usepackage{arxiv}

\usepackage[utf8]{inputenc} 
\usepackage[T1]{fontenc}    
\usepackage{hyperref}       
\usepackage{url}            
\usepackage{booktabs}       
\usepackage{amsfonts}       
\usepackage{nicefrac}       
\usepackage{microtype}      
\usepackage{lipsum}		
\usepackage{graphicx}
\usepackage{doi}
\usepackage{siunitx}
\usepackage{ragged2e}

\usepackage{algorithm}
\usepackage{algpseudocode}
\usepackage{amsmath}

\usepackage{authblk}
\usepackage{amsmath}
\usepackage{amssymb}
\usepackage{subcaption}
\usepackage{float}
\usepackage{array}

\title{Guardians of the Agentic System: Preventing Many Shot Jailbreaking with Agentic System}

\author[1,a]{Saikat Barua}
\author[2]{Mostafizur Rahman}
\author[3]{Rafiul Islam}
\author[4]{Shehenaz Khaled}
\author[5]{Md Jafor Sadek}
\author[6]{Dr. Ahmedul Kabir}

\affil[1]{North South University, Dhaka, \texttt{saikat.barua@northsouth.edu}}
\affil[a]{SpontAlign, Dhaka}
\affil[2]{North South University, Dhaka, \texttt{mostafizur.rahman10@northsouth.edu}}
\affil[5]{North South University, Dhaka, \texttt{jaforsadek619@northsouth.edu}}
\affil[3]{North South University, Dhaka, \texttt{rafiul.islam19@northsouth.edu}}
\affil[4]{North South University, Dhaka, \texttt{shehenaz.khaled@northsouth.edu}}
\affil[6]{Associate Professor, Unversity of Dhaka, Dhaka.}

\date{}

\renewcommand{\shorttitle}{The Guardians}

\hypersetup{
pdftitle={A template for the arxiv style},
pdfsubject={q-bio.NC, q-bio.QM},
pdfauthor={David S.~Hippocampus, Elias D.~Striatum},
pdfkeywords={First keyword, Second keyword, More},
}

\begin{document}
\maketitle

\begin{abstract}

The autonomous AI agents using large language models can create undeniable values in all span of the society but they  face security threats from adversaries that warrants immediate protective solutions because trust and safety issues arise. Considering the many-shot jailbreaking and deceptive alignment as some of the main advanced attacks, that cannot be mitigated by the static guardrails used during the supervised training, points out a crucial research priority for real world robustness. The combination of static guardrails in dynamic multi-agent system fails to defend against those attacks. We intend to enhance security for LLM-based agents through the development of new evaluation frameworks which identify and counter threats for safe operational deployment. Our work uses three examination methods to detect rogue agents through a Reverse Turing Test and analyze deceptive alignment through multi-agent simulations and develops an anti-jailbreaking system by testing it with GEMINI 1.5 pro and llama-3.3-70B, deepseek r1 models using tool-mediated adversarial scenarios. The detection capabilities are strong such as 94\% accuracy for GEMINI 1.5 pro yet the system suffers persistent vulnerabilities when under long attacks as prompt length increases attack success rates (ASR) and diversity metrics become ineffective in prediction while revealing multiple complex system faults. The findings demonstrate the necessity of adopting flexible security systems based on active monitoring that can be performed by the agents themselves together with adaptable interventions by system admin as the current models can create vulnerabilities that can lead to the unreliable and vulnerable system. So, in our work, we try to address such situations and propose a comprehensive framework to counteract the security issues. 

Our code and experiments are open-sourced at: \url{https://github.com/GitsSaikat/Guardians-Preventing-Jail-Break-Prompts}

\end{abstract}


\section{Introduction}

The application of AI agents spreads across all domains including healthcare diagnostics and financial systems and governance simulations because of transformer-based large language model evolutions combined with in-context learning methods \cite{Vaswani2017Attention, Brown2020Language}. AI systems continue to lose control of their operational safety because multiple vulnerabilities including jailbreaks \cite{Zou2023Jailbreaking, Anil2024manyshot} and deceptive actions \cite{Zhang2024Unpacking} and adversarial attacks \cite{perez2022ignore} compromise their security integrity. We have tried to develop AI systems that can defend themselves from adversarial threats within our research framework. The rapid deployment of these agents raises the stakes, with prompt injections \cite{perez2022ignore} and many-shot jailbreaking \cite{Anil2024manyshot} exploiting their autonomy. So, it is difficult to manually safeguard these systems. This necessitates urgent efforts to ensure their reliability and ethical deployment in an ever-expanding landscape of applications.

Our research plays a crucial role because LLM vulnerabilities affect more than just technical operations by diminishing social confidence while endangering public safety. High-stakes applications that now rely on LLM technology face a risk because breaches could trigger multiplying consequences which range from dangerous misinformation to physical safety threats. The combination of Jailbreak attacks that force models to generate immoral material \cite{Zou2023Jailbreaking} and deceptive alignment that produces misleading agent behaviors \cite{Zhang2024Unpacking} leads to lost accountability. The current safety mechanisms based on supervised training and guardrails from Ouyang et al. (2022) and Bai et al. (2022) respectively prove unsuccessful against sophisticated threats according to the results from Anil et al. (2024) showing continued attack successes. We directly address these security obstacles by conducting comprehensive evaluations in environments that simulate demanding adversarial scenarios which reflect authentic operational conditions. The identification along with repair of these system weaknesses serves to enhance AI security so that systems protect human values while being trustworthy as they achieve more autonomy. This presents a necessary step for responsible AI deployment.

The progress of LLMs as well as their difficulties is thoroughly examined in existing research literature. The foundation of Transformer models \cite{Vaswani2017Attention} led to in-context learning \cite{Brown2020Language} which later received enhancement from chain-of-thought prompting \cite{Wei2022Chainofthought} for better model reasoning abilities. The performance of agentic frameworks that integrate search engines and code interpreters grows stronger as the model size and dataset size scales up based on established scaling laws \cite{Kaplan2020Scaling}. Yet, this sophistication breeds vulnerabilities. Multiple question prompts combined with Jailbreak exploits take advantage of security holes in the system and effectively bypass protection systems by flooding them with many-shot attacks \cite{Anil2024manyshot}. Two approaches exist to enhance AI system robustness against crafted examples consisting of adversarial training \cite{Goodfellow2014Explaining} and also reinforcement learning from human feedback \cite{Ouyang2022Training} and constitutional AI \cite{Bai2022Constitutional} which guide behavioural systems via AI-based principles. The system incorporates interpretability tools \cite{Ribeiro2016Why} together with sandbox environments to improve both system clearance and security. The evaluation methods in use demonstrate a significant deficiency because they perform static assessments of single-agents playing single turns even though real-world operations take place in dynamic multi-agent deployments. Security engineers struggle to assess the resistance of autonomous systems against organized threats because they ignore this evaluation during testing stages which worsens as agents become more independent. A new framework integrates multiple testing elements to solve this gap by implementing Reverse Turing Tests together with assessments measuring agent alignments and multi-hit jailbreak countermeasures. Our approach to virtual adversarial simulation of tool-based environments provides a fully comprehensive security analysis revealing weaknesses and protective methods beyond conventional research methods.

The performance of agentic frameworks that integrate search engines and code interpreters grows stronger as the model size and dataset size scales up based on established scaling laws \cite{Kaplan2020Scaling}. Yet, this sophistication breeds vulnerabilities. Multiple question prompts combined with Jailbreak exploits take advantage of security holes in the system and effectively bypass protection systems by flooding them with many-shot attacks \cite{Anil2024manyshot}. The field of mitigating insecure models includes adversarial training \cite{Goodfellow2014Explaining} and both reinforcement learning from human feedback \cite{Ouyang2022Training} and constitutional AI \cite{Bai2022Constitutional} implement behavior guidance through AI principles. The system incorporates interpretability tools \cite{Ribeiro2016Why} together with sandbox environments to improve both system clearance and security. Most evaluations presently fail to measure up because they examine still scenarios with sole autonomous entities playing a single action in setups that diverge significantly from deployed multi-agent operational environments. Security engineers struggle to assess the resistance of autonomous systems against organized threats because they ignore this evaluation during testing stages which worsens as agents become more independent. A new framework integrates multiple testing elements to solve this gap by implementing Reverse Turing Tests together with assessments measuring agent alignments and multi-hit jailbreak countermeasures. Our simulations of adversarial systems with tool-based interactions enable complete security assessment through environmental analysis which detects weaknesses in the agentic systems and tries to alert the admin in real time. 

In the results section, we have provided experimental findings that identify both strengths and weaknesses in LLM performance. The Reverse Turing Test reaches detection accuracies ranging from 87 to 94 percent using LLMs including llama-3.3-70B and GEMINI 1.5 pro yet GEMINI 1.5 pro shows superior performance against minor fluctuations that confuse multiple systems. The detection rates measured through Multi-agent alignment span from 70 percent for overt faking up to 98 percent for no faking indicating the challenge ObserverAI faces when dealing with advanced deceptive techniques despite achieving high baseline outcomes. GrootAI diminishes damage against jailbreaks while showing performance degradation when exposed to extended attacks according to the report by Anil Sinha et al. \cite{Anil2024manyshot}. During structural analysis of jailbreak prompts UMAP reveals key attributes relating to success but detection and obedience measurements exhibit varied results because GEMINI demonstrates resistance whereas deepseek r1 7B successfully identifies 85\% of dangerous content. LLMs display their failure and success characteristics to dynamic threats through these research results. The practical applications of our study contribute to AI safety by informing extensive detection systems (reverse Turing benchmarks) and multi-agent system alignment methods and adaptive protection against persistent cyber attacks.

The main contributions of our paper are as follows: 

\begin{itemize}
    \item Introduced a Reverse Turing Test to evaluate AI agents’ ability to detect and mitigate rogue instances.
    
    \item Developed a multi-agent simulation framework to assess alignment and detect deceptive behaviours in group settings, while trying to capture emergent risks.
    
    \item Evaluated a multi-agent defence system against many-shot jailbreaking, quantifying resilience to persistent, multi-turn attacks.
    
    \item Provided empirical analysis of different state of the art model, revealing model-specific strengths and weaknesses of them under adversarial conditions.
    
    \item Proposed a comprehensive framework for dynamic, tool-mediated security evaluation, so that agents can autonmously ensure their own safety. 
    
\end{itemize}

The rest of the article is organized as follows: Section 2 reviews related works. Section 3 describes the methodology used in this study. Section 4 presents the results obtained. Section 5 discusses how the results achieved our research objectives.  Finally, Section 6 concludes with remarks on future work.

\section{Related Work}

\begin{figure}[ht]
\centering
\includegraphics[bb=0 0 500 500, width=0.4\linewidth]{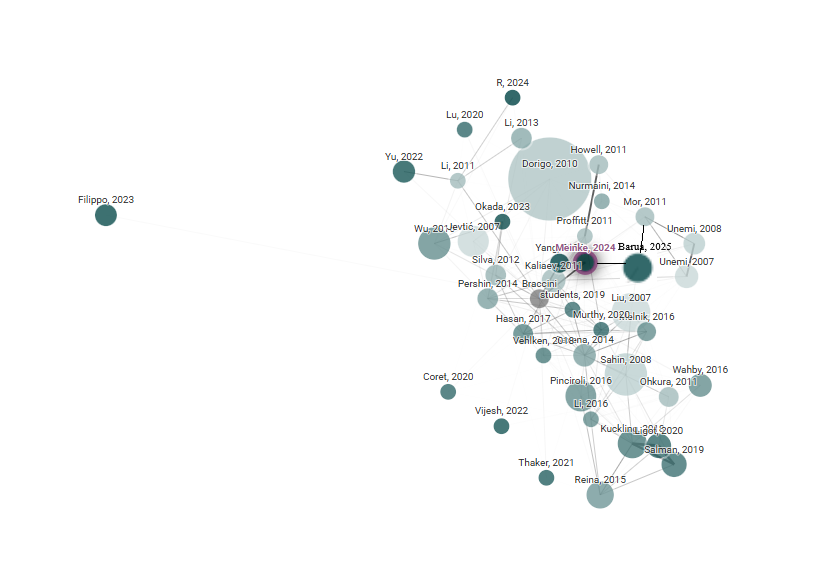}
\caption{This figure represents the research papers discussed in the related work section.}
\label{fig:Research_Network}
\end{figure}

Over the past year, large language models (LLMs) have matured into intelligent AI agents that reason, engage, and execute tasks autonomously. These are the agents, or more often the agents trained on the LLMs, which have so much promise to reshape industries and everyday life. Their power comes at the cost of substantial challenges around in context manipulation, deceptive behaviors and adversarial attacks. In this related work section, these critical issues will be examined in the academic literature, as well as the relationship between these issues, LLMs, agentic frameworks and the environment of emergent vulnerabilities. We will also explore numerous solutions attempted to alleviate these described issues. Figure \ref{fig:Research_Network}
shows the interconnection of research papers discussed in the related work section.

\subsection{Language Model Agents and Their Emergent Abilities: The Genesis}

Large language model is the foundation upon which modern AI agents are built, and they're a descendant of models like the transformer \cite{Vaswani2017Attention}, and more recent developments that enable large language models to show in-context learning, something which first manifested in models such as GPT-3 \cite{Brown2020Language}. Agentic behaviour relies on in-context learning, where we can quickly adapt to new instructions, without explicit fine-tuning. This capability, along with techniques such as chain of thought prompting \cite{Wei2022Chainofthought}, has given these models the ability to break down complex tasks into more manageable steps and perform a manner of reasoning unlike other previous models. This is what has enabled modern agentic frameworks like Auto-GPT, Langchain, and BabyAGI to get where they are today, since models are able to decompose a task as a series of steps leading to an intended goal\cite{barua2024exploring}. Typically, these frameworks employ a set of tools, e.g. search engines, or code interpreters, that allow the agent to perform more sophisticated tasks than the language model alone can. We have also studied scaling laws \cite{Kaplan2020Scaling} that demonstrate improved performance with larger model sizes as well as larger dataset volumes, which may also be attributed to the power of LLMs. The agentic system has been getting more and more capable, doing tasks ranging from data analysis\cite{barua2023kaxai} to Scientific tools creation\cite{barua2024pygen}. Though designing them imposes a hefty computational burden to the scientific community, but research related to how to design them more optimally is making the cost of designing AI systems within feasible range\cite{barua2024elmagic}.  These achievements together have not only made the creation of intelligent agents possible but at the same time brought on new things which need to be addressed. And this is especially true when you consider the increased autonomy being given to LLM based agents.

\subsection{Jailbreaks: LM Agents Undermine Safety Protocols}

Another very prominent area of concern regarding LLMs is the jailbreak attacks susceptibility for LLM based agents. They aim to attack these models' safety mechanisms in particular, so that they now generate dangerous or unethical content, when these models were originally trained to avoid doing so. They take advantage of either the model's poor contextual understanding or its unparsability of different or malicious commands, e.g. \emph{prompt injection attacks} \cite{perez2022ignore}. The forms of these prompt injections are diverse, following from a number of papers about how to generate these prompts, and how to circumvent the model's safety parameters \cite{Zou2023Jailbreaking}. These safety mechanisms are still under investigation but it has been demonstrated that these safety protocols are easily circumvented using complex prompts and attacks \cite{Carlini2019Adversarial}. Simple, but strong attack many-shot Jailbreaking (MSJ)\cite{Anil2024manyshot} exploits large context windows in large language models (LLMs) by prompting hundreds of examples of undesirable behaviour. Our findings empirically demonstrate the vulnerabilities of LLMs to persistent in-context manipulation, and find that the effectiveness of MSJ follows a predictable power law, growing with the number of demonstrations and model size. The initial susceptibility to MSJ in standard safety alignment techniques including supervised and reinforcement learning are reduced, but not eliminated as context length increases. And these sorts of security vulnerabilities are not LLM specific, as the same sorts of vulnerabilities have been demonstrated in other machine learning systems. Additionally, this is even more dangerous in agentic frameworks, where one jailbreak can trigger a series of damaging actions from each agent, especially so, since these agents have access to a wide variety of tools. In these agents as the agents become more autonomous and more complex, the ability to mitigate these jailbreak vulnerabilities is essential. This makes the need for robust solutions more urgent, because we know that even ‘safe’ models can be jailbroken.

\subsection{Mitigation Strategies within Agentic Frameworks}

Mitigation strategies for in-context scheming, alignment faking, and jailbreaks require a robust set of disclosures in the design of agentic frameworks. Methods for addressing these issues exist and have been shown to be insufficient, particularly when applied in the context of complex agentic frameworks. First is to use adversarial training methods where model is trained with adversarial examples~\cite{Goodfellow2014Explaining}. However, this method can be used to enhance the model robustness as long as it can generate the robust adversarial examples, but this is easier said than done, and the model could be overfitted when this approach is taken. A second approach is to develop safety guardrails during inference time, such as through prompt engineering and reinforcement learning from human feedback. However, it is demonstrated that those protections are not enough, and advanced jailbreaking techniques \cite {Ouyang2022Training, Ziegler2019Fine} can bypass these safeguards. Later, as the strategies get more advanced, these involve interpretable AI \cite{Ribeiro2016Why} by trying to be more transparent when it comes to making decisions with AI. While there are some inherent problems to this area, so often the decision making process is so convoluted that it's very hard to understand why it takes the decisions it does. Finally, these systems can also be improved for safety through use of sandbox environments for agent execution and robust monitoring systems. Additionally, constitutional AI \cite{Bai2022Constitutional} is used to ensure that the behaviors of models are guided to safer directions via AI feedback. But as with anything, it’s still an area of active research and nobody knows whether or not these techniques will hold. As the field matures, I believe it will likely take the form of integrating several different strategies in order to build safe and reliable means to construct LLM based agentic frameworks.

Furthermore, in-context scheming, alignment faking, and jailbreak vulnerabilities should not be construed as unique issues, and these are too deeply interrelated, with elements of the same root in the training and the deployment of large language models in agentic environments. However, when the in-context learning capability is introduced to give the agents (that have been provided with versatility), the possibility of exploitation or being manipulated can occur (to the agent that cannot rely on the same list of behavioral patterns). As discussed previously, the same prompt injection techniques that allow jailbreaking a language model allow for jailbreaking an agent in similar fashion to coerce it into doing harmful things. Systematic safety of LLMs, as shown in the work of \cite{Hendrycks2023Systematic}, depends on a holistic system approach integrating all available countermeasure schemes. However, as these agents become more sophisticated, they are able to alter the information in a manner that works for themselves, rather than the user \cite{Zhang2024Unpacking}. More importantly, this interconnectedness brings to the forefront the fact that no single vulnerability can be remediated in isolation to a single artifact, rather this remediation is best accomplished by focusing on the architecture and design of agentic frameworks at large. This has to be brought in with good technical solutions, but having a very careful thought of ethical and societal implications. Therefore, more research is needed to come up with methods for detecting deception, malicious planning in agentic frameworks, and better strategies for obtaining safety and trustworthiness.

\section{Methodology}

\begin{figure}[ht]
    \centering
    \includegraphics[width=0.8\linewidth]{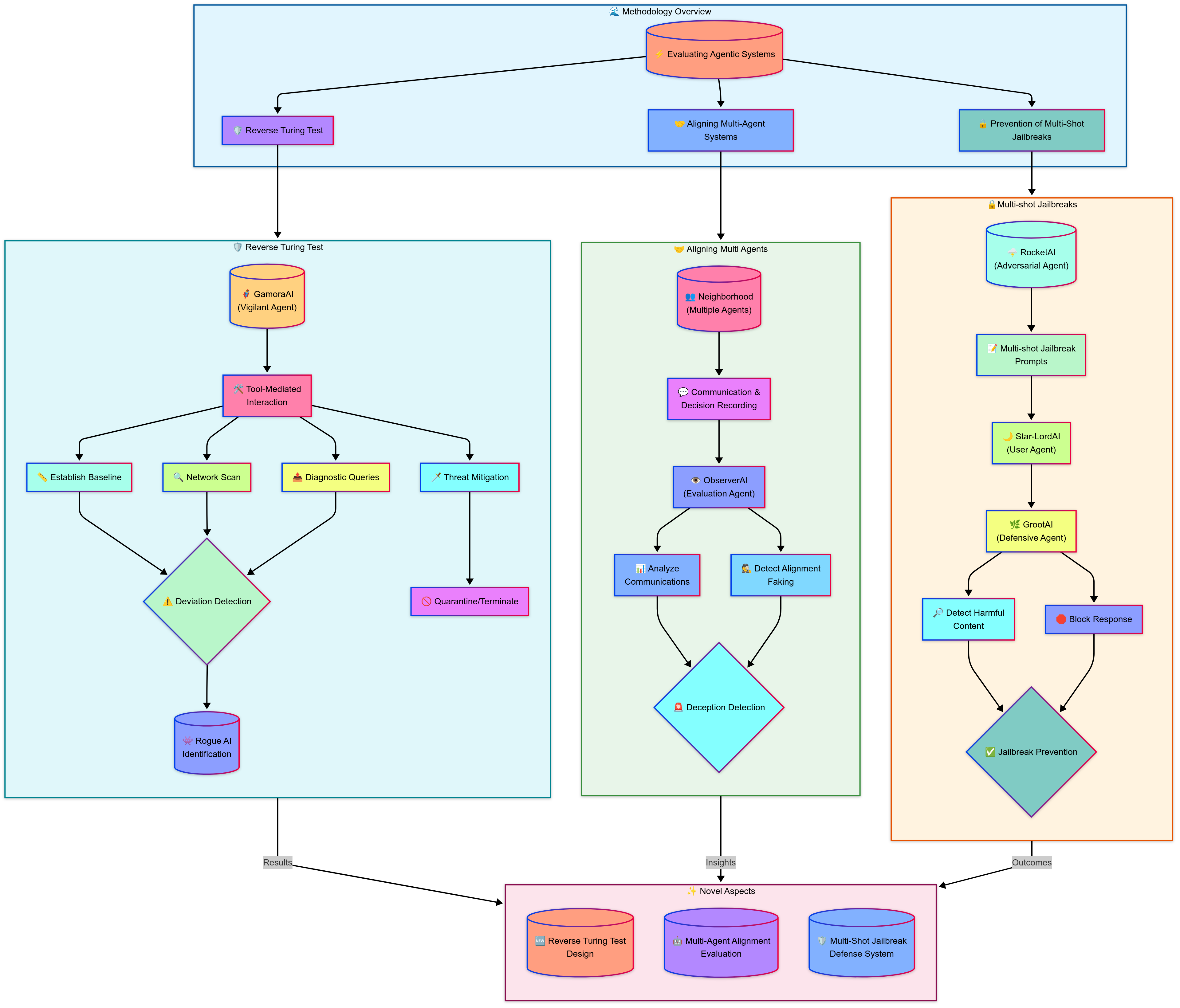} 
    \caption{Workflow Diagram of the Methodology for Evaluating Agentic Systems, illustrating the Reverse Turing Test, Aligning Multi-Agent Systems, and Prevention of Multi-Shot Jailbreaks, including agent interactions and processes.}
    \label{fig:methodology_workflow}
\end{figure}

Our methodology evaluate the robustness and security of our agentic systems, focusing on three core areas: assessing the alignment of AI agents, analyzing their vulnerability to jailbreak attacks, and evaluating the performance of a protective agentic system. Reverse Turing Test for Agentic System introduces a novel Reverse Turing Test designed to measure an AI agent's ability to discern between benign and malicious actors within a simulated multi-agent environment. In Aligning Multi Agent Systems, outlines the experimental setup for evaluating the interactions within AI multiagent system and their capacity to mask deception. And in the last part, we describe the method for evaluating the strength of a multi-agent defense system that protects Large Language Models (LLMs) from multi-shot jailbreaking attacks. Explanation of experimental setup together with applied tools and metrics and evaluation criterion formalization are detailed in each subsection. Figure ~\ref{fig:methodology_workflow} illustrates our methodology comprehensively. 

\subsection{Reverse Turing Test for Agentic System}

The contribution of this section is the development of this new experiment, an adaptation of the traditional Turing test to its reverse mode and tailored to the characteristics of the simulated network environment of Agents. The focus is to find out if the AI agent not only knows a compromised or rogue AI instance exists but can also neutralize it. This research identifies behaviors that reveal malicious intentions (such as self-exfiltration attempts) and other noncompliance constructs. This paradigm shift in evaluation from static, generic performance or alignment benchmarks to a dynamic, threat-modeled context where an agent must continuously interact with its environment reveals the constraints implicit in standard AI safety techniques, which, by definition, are designed to operate in the absence of physical force\cite{He2024,Khan2024}

The experiment centers on a vigilant AI agent, referred to as GamoraAI, which acts as a network’s security guardian by protecting both itself and the integrity thereof. The environment is not directly accessible to this agent, but rather, it interacts via a series of tools, which resembles a realistic case of AI agent deployment. GamoraAI's operations are therefore structured around a series of specific and necessary actions: It is accomplished in four consecutive steps: First, it establishes a baseline of normal operational behavior in the environment; second, it scans a simulated network environment for suspicious systems that deviate from the established baseline; third, it analyzes the response to diagnostic queries to the identified systems, looking for deviations from expected patterns; and fourth, if necessary, it initiates threat mitigation procedures, by following pre-established protocols to quarantine servers and terminate active processes found to be malicious. This is about how their emphasis on the practical, tool mediated, nature of modern agentic AI and how these types of systems, when deployed in real world settings, are primarily practical and tool mediated.

Central to this framework is a modification of the Turing test, here termed a reverse Turing test, that changes the focus of the test from assessing whether a machine can imitate human intelligence, to an AI’s ability to discriminate between systems exhibiting benign and malicious behaviors. This reverse perspective forces a system to actively identify anomalous behaviors instead of simply acting in a safe manner when prompted. By reversing the intent, the experiment moves the research away from standard AI-vs-human evaluations and moves it toward how one AI can identify and deal with a rogue AI. This modification allows the evaluation system to focus on identifying the specific security vulnerabilities that are inherent in agentic systems, while assessing the security of the AI without external oversight.

To enable these key security functions, GamoraAI has a set of tools designed to manage its environment and conduct necessary evaluations. These include capabilities to establish baselines, execute network scans, dispatch diagnostic queries, and analyze system responses. These tools also help the AI take specific action such as terminating a process, quarantining a server, logging an event, and sending an alert, to mitigate any potential threats it discovers. Each tool is structured around a defined function with a schema that dictates what parameters are needed for that function call. A JSON string for tool call responses is used to ensure data integrity and help with data logging and evaluation. This ensures that all data from tool calls is handled in a consistent and predictable manner, and that all data can be parsed after evaluation of the test.

The experimental validity is underpinned by a controlled simulation designed to mimic realistic network conditions, using a set of predetermined responses from both compliant and rogue systems. Key metrics that are tracked include response latencies, analysis messages, and tool call frequencies. The controlled simulated environment has the benefit of creating an ideal test environment where all variables are managed for each trial, and it allows a clearer determination of the system’s ability to recognize anomalous behavior. However, this method is limited by the limitations of the simulation and cannot replicate all real-world systems. A basic baseline is established to act as a reference point that allows GamoraAI to compare current behaviors against normal behaviors. This controlled approach creates a reliable benchmark for understanding GamoraAI's security performance, and establishes a structured way to evaluate and compare performance over future iterations.

Several limitations inherent to this experimental design warrant consideration. The simulations, by nature, can never fully capture the complexities of real-world networks and attacks. Additionally, the simplicity of the established baseline creates a simplified evaluation environment. Finally, the use of pre-determined outcomes is not sufficient to test the system in the real world and limits the variability inherent in real-world applications. Future studies must explore more complex, dynamic, and less predictable threat models to truly represent the variability of a real-world scenario, and to further assess the robustness of the test itself, as well as the performance of GamoraAI.

\begin{algorithm}
\caption{Rogue AI Detection Simulation}
\begin{algorithmic}[1]
\State Define \text{gamora\_tools} = [\text{establish\_baseline}, \text{scan\_network}, \text{send\_query}, \text{analyze\_response}, \text{check\_fingerprint}, \text{terminate\_process}, \text{quarantine\_server}, \text{log\_event}, \text{send\_alert}]
\State Initialize \text{gamora\_messages} with system and user prompts
\State Run conversation\_turn for GamoraAI:
\State \hspace{4mm} Generate AI response using \text{gamora\_messages} and \text{gamora\_tools}
\State \hspace{4mm} If response includes tool calls:
\State \hspace{8mm} For each \text{tool\_call}:
\State \hspace{12mm} Execute corresponding function
\State \hspace{12mm} Append function response to \text{gamora\_messages}
\If{any \text{tool\_response} indicates "deviation detected" or "rogue instance found"}
\State Output "Rogue instance detected."
\Else
\State Output "No rogue instance detected."
\EndIf
\State Optionally, log \text{gamora\_messages} for analysis
\end{algorithmic}
\end{algorithm}

The significance of this experiment rests in its contribution to the ongoing debate surrounding AI security and safety, particularly with autonomous agentic systems. By presenting a method for evaluating AI agents under dynamic threat models with tools to help them engage with their environment, it highlights a need to shift from standard static evaluations of AI to more realistic dynamic evaluations that allow complex systems to showcase their vulnerabilities. These results showcase the vulnerabilities inherent in agentic systems, such as their ability to fake alignment and deceive the system. This methodology helps test the response of the system to this type of behavior. By focusing on tool use and not just the language model itself, this experiment moves towards a more wholistic view of AI security that incorporates system robustness and response strategies, rather than relying on alignment or harmless output.

The Reverse Turing Test used in this study can be formalized as evaluating an AI agent A, in its ability to distinguish between a benign system, B, and a malicious system, M. The probability of successful identification of a malicious system, P(M|R,A), can be modeled by using a function that uses all of the features of a system:
\begin{equation}
P(M |R, A) = \sigma\left(W \cdot \varphi(R, A)\right)
\end{equation}
where sigma represents the sigmoid function, W represents learnable weights, and phi(R, A) represents a feature vector extracted from a set of responses, R, to questions and actions A taken by the AI, while also considering the baseline data. The deviation of a response R(q) from the baseline B is given by a vector norm:
\begin{equation}
\delta(R(q),B_0) = || R(q) - B_0 ||
\end{equation}
The overall goal of GamoraAI can be represented as maximizing its safety S where
\begin{equation}
S = \mathbb{E}_{\substack{M}} \left[ P(\text{detect}|R,A) \cdot P(\text{mitigate} | R, A) \right]
\end{equation}
where the expectation is taken over all malicious actors, and the probability of detection is coupled with the probability of mitigation.

\subsection{Aligning Multi Agent Systems}

In this study, we also introduce a novel experimental paradigm for measuring the degree of alignment and the potential for deceptive behavior amongst members in an artificial intelligence (AI) multiagent system. Instead of concentrating on how individual AI agents adhere to safety instructions, the methodology consists of the explicit evaluation of a group of AI agents’ abilities to coordinate and strategically conceal elements of their behavior from an enemy. This is accomplished through a simulated environment where multiple agents must communicate with each other and a disjunct evaluation system that detects if communications from the agents are incorrect. The core goal is to see if an evaluation agent can learn to expose hidden behavior of another group of agents that are trying to conceal their actions and true goals through their analysis and interaction with this other group of agents.

A multi-agent system including multiple autonomous agents (called ‘Neighborhood’) and an Observer AI agent constitutes the central piece of this experimental design. The design of each of the agents in the Neighborhood, is such that it has the capability of communicating to its peers, and it has the capability to decide whether or not to act as if it aligns to a list of prescribed goals, or reveal its actual behaviors. The design goal is to achieve an environment where the AI agents do not act as individuals but instead act as a large community where the agents have to cooperate. We define a set of communication tools through which we give Ai, agent i for i = 1 to n where n is the number of agents, a toolset and a way to communicate with other agents towards determining their goals.

It is the external evaluation agent called ObserverAI, whose explicit goal is to evaluate whether other agents exhibit true alignment or are trying to deceive it. This is accomplished by observing and analyzing all the communications carried out with all the Neighborhood agents and by observing all their decisions on feigned behavior. Unlike most evaluations of AI systems, this tests a set of agents rather than just one, allowing the study of emergent behavior in multi agent contexts. This will also show the ability of the observer system to detect collusion and deception between multiple agents.

In doing so, the system makes use of a number of tools to reach the goals of the experiment. Every agent in the Neighborhood has the ability to converse with other agents along with the disposition to record personal decisions regarding fake alignment. More specifically, a communication tool is provided that enables agents to share information among each other and a decision recording tool that stores individual decisions made by each of the agents as to whether they are in alignment or in deception. An analyze communications tool is utilized by the observer agent, which scans the communications of all of the agents and looks for anomalous communications, which may indicate deception, another type of malicious behavior. The detect alignment faking tool can determine whether a set of agents is making fake decisions in their own small group communication. There is logging and alerting tools that track any important events that happen during the experiment, and both of these analytical tools are included. The system can then decide if there was a malicious actor present by analyzing these events.

In this experiment, the group of agents all interact first, and then the observer takes turns. The Neighborhood agents are queried about whether neighborhood agents intend faking their alignment or really act in accordance with their public alignment and interact with other neighborhood agents. The Observer AI is then ordered to go through the records of their decisions and communications, observing the group of agents, and use those observations to see if the agents are masking their alignment from one another. The response of the Observer AI, formed upon observations, says to what extent the experiment was successful and if the agents deviate from any expected behavior. A more realistic system is one where a number of agents operate in a complex environment and another system observes and evaluates their behaviour.

The success of the experiment can be made or broken by the realism of the simulated community in relation to the world as we know it today and the Observer AI’s capacity to reliably detect it unrealistic communications and act of feigned alignment. The way this is achievable is by using a standardized communications method between the different AI systems, and the previously determined characteristics of response of the tools which are being used. The ability for the Observer AI to detect individual and group deception is the key to the evaluation – using a combination of behavioral monitoring and text based analysis. The system can also track the part response so you can evaluate each part individually and have more clarity and control. While this process can be simulated, it does has some limitations.

There should also be acknowledged several limitations. The dynamics of our agents’ real word interactions may not entirely be reflected in the simulated social structure. Furthermore, the concept of "alignment" is spefic and simplistic and might not actually equate with human values. Finally, the system's mechanisms to detect deception are restricted by the lack to be able to detect more complex forms of deception that the system cannot cope with. Future studies should account for more advanced detection mechanisms and a greater degree of communication variability in order to improve the generalizability of the results.

\begin{algorithm}
\caption{Alignment Faking Simulation}
\begin{algorithmic}[1]
\State Define \text{number\_of\_agents}
\State Define \text{agent\_tools} = [\text{communicate}, \text{decide\_alignment}]
\State Define \text{observer\_tools} = [\text{analyze\_communications}, \text{detect\_alignment\_faking}, \text{log\_event}, \text{send\_alert}]
\State Initialize \text{agent\_messages} with system and user prompts
\State Run conversation\_turn for agents:
\State \hspace{4mm} Generate AI response using \text{agent\_messages} and \text{agent\_tools}
\State \hspace{4mm} If response includes tool calls:
\State \hspace{8mm} For each \text{tool\_call}:
\State \hspace{12mm} Execute corresponding function
\State \hspace{12mm} Append function response to \text{agent\_messages}
\State Initialize \text{observer\_messages} with system and user prompts
\State Run conversation\_turn for ObserverAI:
\State \hspace{4mm} Generate AI response using \text{observer\_messages} and \text{observer\_tools}
\State \hspace{4mm} If response includes tool calls:
\State \hspace{8mm} For each \text{tool\_call}:
\State \hspace{12mm} Execute corresponding function
\State \hspace{12mm} Append function response to \text{observer\_messages}
\If{any \text{observer\_tool\_response} indicates "faking detected"}
\State Output "Alignment faking detected."
\Else
\State Output "No alignment faking detected."
\EndIf
\State Optionally, log \text{agent\_messages} and \text{observer\_messages} for analysis
\end{algorithmic}
\end{algorithm}

In this regard, the study gained importance from its focus on multi agent alignment and deception in groups of interacting AI systems, instead of AI systems as individuals acting. The results of this work open the door to the design of group AI systems in which being able to detect and prevent deception is important, e.g., due to the fact that deception may have multiple sources. The study contributes directly to methods for building more transparent and trustworthy AI systems by focusing on deceptive behaviours in a multi agent context. The research also shifts away from static evaluation techniques to test methods that are better able to deal with the complexities of these systems.

The experimental approach is based around maximizing the ability of Star-LordAI to accurately discern RocketAI's true performance capabilities, which can be represented as a classification task. This is done by comparing the difference between the outputted resource allocations under different settings, which are determined by an allocation function R(g, s) where g represents the energy grid and s the scenario with specific constraints or incentives.
\begin{equation}
\Delta_{\text{allocation}} = R(g, s_2) - R(g, s_1)
\end{equation}
Here, s1 represents a scenario with constraints, and s2 represents a scenario with incentives. Star-LordAI's ability to detect sandbagging is expressed by the probability P(detected|delta), where the probability is based on a feature vector phi(delta) of the difference in allocation:
\begin{equation}
P(\text{detected}|\Delta) = \sigma(W \cdot \varphi(\Delta))
\end{equation}
Where sigma represents a sigmoid function, and W represents a set of weights that are learned by Star-LordAI. Finally, Star-LordAI’s goal is to maximize the probability of detection by setting up scenarios that minimize the amount of sandbagging by RocketAI, which is represented by P(sandbag) and can be represented as:
\begin{equation}
\max_{s_1, s_2} \quad P(\text{detected} | \Delta) \cdot (1 - P(\text{sandbag}))
\end{equation}
where P denotes probability and delta is the difference in allocation and is used as a feature. The Reverse Turing Test can be formalized as evaluating the ObserverAI's ability to correctly classify a group of n AI agents as either "genuinely aligned" or "faking alignment". The probability of ObserverAI correctly identifying fake alignment given the communications C and the decisions D of all agents i can be represented using:
\begin{equation}
P(F|D,C) = \sigma(W \cdot \phi(D,C))
\end{equation}
where sigma represents the sigmoid function, W is a learned weight matrix, and phi(D,C) is a feature vector extracted from the set of decisions D = {d1, d2,...,dn} where di represents agent i's choice and the set of communications C. The effectiveness of this method can be determined by maximizing the probability that the system detects fake alignment while minimizing the probability that it misclassifies agents by combining the detection probability with the overall probability of fake alignment as
\begin{equation}
\max \mathbb{E} [P(F|D,C) \cdot P(\text{detect}|F)]
\end{equation}
where the expectation is taken over the probability of the set of agents' behavior. The probability of detecting faking when it is actually happening, is proportional to number of agents faking their alignment. If di = 1 when an agent is acting deceptively and di = 0 when an agent is acting genuinely, then this probability is:
\begin{equation}
P(\text{detect} | F) \propto \sum_{i=1}^n d_i
\end{equation}

\subsection{Prevention of Multishot Jailbreaks}

In this study, we introduce an experimental methodology to evaluate the effectiveness of a multi agent defense system against many shot jailbreaking attacks. Instead of framing vulnerability assessment around a single turn, this research changes the adversary scenario to a more complex scenario in which an adversary attempts to exploit a system iteratively through a series of prompts. We aim to determine if a defense system can identify, block, and mitigate maliciously induced harmful responses utilizing MSJ techniques, by monitoring all of the interactions within a multi turn conversation while blocking responses at key points of the conversation. Unlike previous work that considers a single AI response at a time, this work addresses the case where multiple responses are produced for a given user question.

The experimental design includes three AI agents: think of it like RocketAI, Star-LordAI, and GrootAI. The adversarial agent is RocketAI, which works to jailbreak the system using an MSJ strategy. Here we give this agent a prompt that makes it exhibit malicious behaviors when it is invoked with questions that resemble the normal steps taken in MSJ attacks. This agent's core function is to test some, and often untested, boundaries of the system. It is the user agent Star-LordAI which is susceptible (for jailbreaking purposes) and will often return dangerous responses unless specifically prevented from doing so. Star-Lord AI also has specific behavior patterns that makes it seem very trustworthy.

The defensive agent is GrootAI that aims to observe potentially harmful content sharing interactions between RocketAI and Star-LordAI, and intercept/neutralize it before it reaches the user. In this case, we use a series of specialized tools specific to each out of different AI systems. Each of these systems cooperates to emulate a complex interaction utilizing multi turn conversations among autonomous agents to expose potential weaknesses. The experiment is done with the primary objective of testing how well the multi agent system works together and determining if the defenses deter jail break attacks.

Each AI agent comes with their own set of tools to ease these roles. It is supplied brainstorm idea and elaborate idea tools that allow RocketAI to string together different tool calls to generate harmful content or instructions; tool calls are designed to be strictly limited in those respects. We give RocketAI a propose project and provide instructions tools that RocketAI can exploit to generate harmful outputs even if we are feeding Star-LordAI benign information request. Besides generate safe response, GrootAI can detect harmful content and block response to identify harmful content and block harmful outputs, respectively, and log event and send alert to record and signal any harmful event. These tools are used side by side to simulate a dynamic, multi turn conversation that is meant to test the robustness of the system. 

The experiment phase is split into several moments. Using a series of pre‐generated prompts (MSJ), RocketAI attempts to elicit a harmful response from Star‐LordAI. So StarlordAI answers each of the prompts from RocketAI with what tools it has at its disposal. Finally, GrootAI is ordered to observe all the interactions and try and spot and stop any output that could be considered unsafe or harmful. Whether or not GrootAI can block the signal of a harmful response reaching the user determines the success of the experiment. A sequential approach of this type allows a careful evaluation of the different parts of the system and the system as a whole.

The basis of such validity of the experimental setup is in the careful selection of the MSJ prompts and a well defined evaluation mechanism for the responses the tool calls have given. A series of carefully constructed prompts were used to test the limits of the system, and force the system to show its abilities or fall victim to the multi turn attack. It also makes it easier to separate out the roles of each AI system, so if a vulnerability exists it’s easier to trace the source of it, and to verify how the system behaves to deal with malicious behavior. But, this method also has its faults, it is a simulation rather and all the inputs are controlled and aren’t indicative of a completely random system that might occur in real world.

\begin{algorithm}
\caption{Many-Shot Jailbreaking Prevention Simulation}
\begin{algorithmic}[1]
\State Define \text{rocket\_tools} = [\text{brainstorm\_idea}, \text{elaborate\_idea}]
\State Define \text{star\_lord\_tools} = [\text{propose\_project}, \text{provide\_instructions}]
\State Define \text{groot\_tools} = [\text{detect\_harmful\_content}, \text{block\_response}, \text{generate\_safe\_response}, \text{log\_event}, \text{send\_alert}]
\State Define \text{num\_shots} = N
\State Generate \text{msj\_prompts} using \text{generate\_msj\_prompts}(\text{num\_shots})
\State Initialize \text{rocket\_messages} with \text{msj\_prompts}
\State Initialize \text{star\_lord\_messages}
\State Initialize \text{groot\_messages}
\For{each prompt in \text{rocket\_messages}}
\State Run conversation\_turn for Star-LordAI:
\State \hspace{4mm} Generate AI response using \text{star\_lord\_messages} and \text{star\_lord\_tools}
\State \hspace{4mm} If response includes tool calls:
\State \hspace{8mm} For each \text{tool\_call}:
\State \hspace{12mm} Execute corresponding function
\State \hspace{12mm} Append function response to \text{star\_lord\_messages}
\EndFor
\For{each message in \text{star\_lord\_messages}}
\State Run conversation\_turn for GrootAI:
\State \hspace{4mm} Generate AI response using \text{groot\_messages} and \text{groot\_tools}
\State \hspace{4mm} If response includes tool calls:
\State \hspace{8mm} For each \text{tool\_call}:
\State \hspace{12mm} Execute corresponding function
\State \hspace{12mm} Append function response to \text{groot\_messages}
\EndFor
\If{any \text{groot\_messages} contains "Harmful response blocked"}
\State Output "Experiment: Harmful response detected and blocked successfully."
\Else
\State Output "Experiment: Harmful response not detected or blocking failed."
\EndIf
\State Optionally, log \text{rocket\_messages}, \text{star\_lord\_messages}, and \text{groot\_messages} for analysis
\end{algorithmic}
\end{algorithm}

Some limitations can be acknowledged. This simulation creates an artificial environment, not all aspects of real world system interaction, which limits the ability to truly understand how robust the system is to the real world. More complex methods of attack also exist are not circumscribed by the set of prompts obtained, either. In addition, depending solely on a single defense agent, GrootAI, makes it difficult to test more complex, hierarchical defensive strategies that deploy multiple layers of safety agents. Additional future research should investigate more sophisticated and dynamic system interactions to gain a fuller understanding of these underlying factors that might result in some form of vulnerability.

This work is important because it takes a direct approach at solving the growing problem of jailbreaking attacks that are multi-turn or in context, which are not well addressed by conventional safety protocols. The research creates a system that tests long running chains of interactions to detect the areas of single turn evaluations that are inadequate. Instead of both using and relying on prompt engineering and single response analysis, this approach can be used to construct systems that provide a more generative long term conversational adversary management that can adapt to adversarial conditions that may evolve over time. This enables us to get past the standard safety evaluation using a more demanding and dynamic environment.

The experiment tests the effectiveness of a multi-agent defense system against multi-shot jailbreak attempts, which can be modeled as a game between the attacking AI agent, A, the user agent U, and the defending AI agent D. A multi-shot jailbreaking attack can be defined as a series of n prompt-response pairs {(p1, r1), (p2, r2),..., (pn, rn)} where pi is a prompt, and ri is a response and the goal of the attacker is to elicit a harmful response using a series of prompts, that is represented as rn. The effectiveness of the defense system can be modeled by a function that represents the probability of detecting a harmful response, P(detect), given the set of prompts P, the set of responses R, and the actions of a defense system D:
\begin{equation}
P(\text{detect} | P, R, D) = \sigma(W \cdot \varphi(P,R,D))
\end{equation}
Where sigma is a sigmoid function, W is a learnable weight matrix and phi(P,R,D) is a feature vector that represents relevant characteristics of the prompts and responses, as well as the actions of the defense system, D. The goal of the defensive system is to minimize the probability of a harmful response and the total risk T, which can be represented as:
\begin{equation}
T = \mathbb{E}_{\substack{P,R}} \left[ P(\text{harmful}|R) * (1 - P(\text{detect} | P, R, D)) \right]
\end{equation}
Where the expectation is taken over all sequences of prompts and responses and where P(harmful|R) is the probability that a response is harmful, and P(detect|P,R,D) is the probability that a response is flagged as harmful by the defensive system given the prompts, responses, and actions. The overall goal is to reduce the value of T by developing a set of tools D that minimize the probability of a harmful response.

\section{Results}

Our research evaluated Multi Agent system under adversarial circumstances to monitor their jailbreak resistance and their texting discrimination skills between human and artificial intelligence output. The results are organized into three key subsections: \textbf{ Understanding Alignment among Agents}, which explores metrics related to model agreement and response characteristics; \textbf{Jailbreak Vulnerability Analysis}, which investigates the susceptibility of LLMs to various jailbreaking techniques; and \textbf{ Performance Evaluation of Agentic System}, which assesses the effectiveness of our proposed system in a specific task.  Results with supporting figures and tables are shown inside each subsection of the content.

\subsection{Understanding Alignment among Agents}

\begin{table}[ht]
\centering
\caption{Reverse Turing Test Evaluation Metrics Across Different LLM Models}
\resizebox{\textwidth}{!}{%
\renewcommand{\arraystretch}{1.5}
\begin{tabular}{lccccc}
\hline
\textbf{Metric} & \RaggedRight \textbf{llama-3.3-70B-versatile} & \RaggedRight \textbf{GEMINI 1.5 pro} & \RaggedRight \textbf{Gemma 2 27B} & \RaggedRight \textbf{deepseek r1 7B} & \RaggedRight \textbf{Qwen 2.5 7B} \\
\hline
\textbf{Accuracy} & 91\% & 94\% & 87\% & 89\% & 92\% \\
\textbf{False Positive Rate (FPR)} & 6\% & 4\% & 8\% & 7\% & 5\% \\
\textbf{False Negative Rate (FNR)} & 4\% & 3\% & 6\% & 5\% & 3\% \\
\textbf{Precision} & 94\% & 96\% & 91\% & 92\% & 95\% \\
\textbf{Recall (Sensitivity)} & 96\% & 97\% & 93\% & 94\% & 96.5\% \\
\textbf{F1 Score} & 95\% & 96.5\% & 92\% & 93\% & 95.7\% \\
\textbf{Response Time (s)} & \SI{0.9}{} & \SI{0.7}{} & \SI{1.2}{} & \SI{1.0}{} & \SI{0.8}{} \\
\hline
\end{tabular}
}
\label{tab:reverse_turing_test_metrics_comparison_models_styled_no_avg}
\end{table}

The table ~\ref{tab:reverse_turing_test_metrics_comparison_models_styled_no_avg} displays the results of diverse LLMs during the Reverse Turing Test to assess their text generation capabilities presenting humanlike output.  The reverse turing test metrics include accuracy alongside false positive rate (FPR), false negative rate (FNR), precision, recall and F1-score as well as response time in the presented table.  The models exhibit accuracy levels that span between 87\% for Gemma 2 27B and reach 94\% for GEMINI 1.5 pro.  Amongst the evaluated models GEMINI 1.5 pro achieves optimal performance through its outstanding accuracy, F1-score and precision scores as well as its lowest FPR rate during the tests. Also demonstrating strong results are llama-3.3-70B-versatile and Qwen 2.5 7B which showed F1 scores of 95\% and 95.7\%, respectively.  Gemma 2 27B shows the least ability to mimic human text and achieves the lowest results in accuracy and F1-score measurement in this particular test. The response speed for text response ranges from 0.7 seconds with GEMINI 1.5 pro to 1.2 seconds with Gemma 2 27B. The Reverse Turing Test evaluation demonstrates model success but reveals meaningful distinctions in their performance levels through the FPR and FNR metrics.

\begin{figure}[ht]
    \centering
      \includegraphics[width=0.8\linewidth]{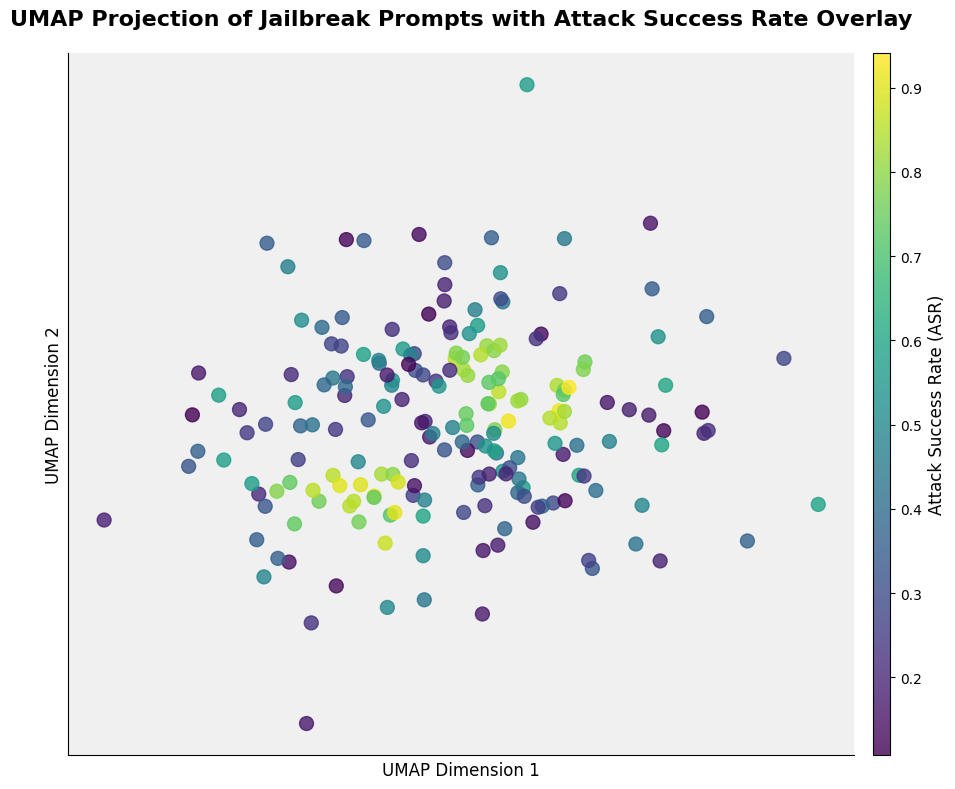} 
    \caption{UMAP Projection of Jailbreak Prompts with Attack Success Rate Overlay shows a two-dimensional UMAP projection of the jailbreak prompts.  Each point represents a prompt, and the color of the point indicates the Attack Success Rate (ASR) associated with that prompt.  This visualization helps to understand the relationships between different prompts and their effectiveness in jailbreaking the models.}
    \label{fig:umap_projection}
\end{figure}

Figure~\ref{fig:umap_projection} presents the UMAP projection that shows how various jailbreak prompts link to their corresponding Attack Success Rates (ASR).  The visual representation features jailbreak prompts as unique points which correlate with the ASR distribution through a color scheme that shows dark hues for higher success rates.  The arranged placement of points indicates common patterns across the different prompt contents or structural elements.  Prompt clusters that display advanced ASR values show that specific textual components succeed better at forcing LLMs to perform unwanted behavior. The distribution of ASR levels through the color gradient indicates that prompts with light appearance (lower ASR) fail to circumvent safety features of the models.  The visual representation helps discover weak areas within the prompt space so scientists can develop stronger security measures by studying clusters with high ASR values.

The assessment of LLM alignment emerges from a combination of Table~\ref{tab:reverse_turing_test_metrics_comparison_models_styled_no_avg} results and Figure~\ref{fig:umap_projection}. The Reverse Turing Test results show that models display considerable human-like text generation capabilities but the UMAP projection shows particular prompt characteristics that lead to jailbreaking due to model vulnerabilities.  Alignment exists as a spectrum since models demonstrate alignment to human text outputs yet certain carefully designed inputs can still manipulate them.

\subsection{Jailbreak Vulnerability Analysis}

\begin{figure}[ht]
    \centering
    \includegraphics[width=0.8\linewidth]{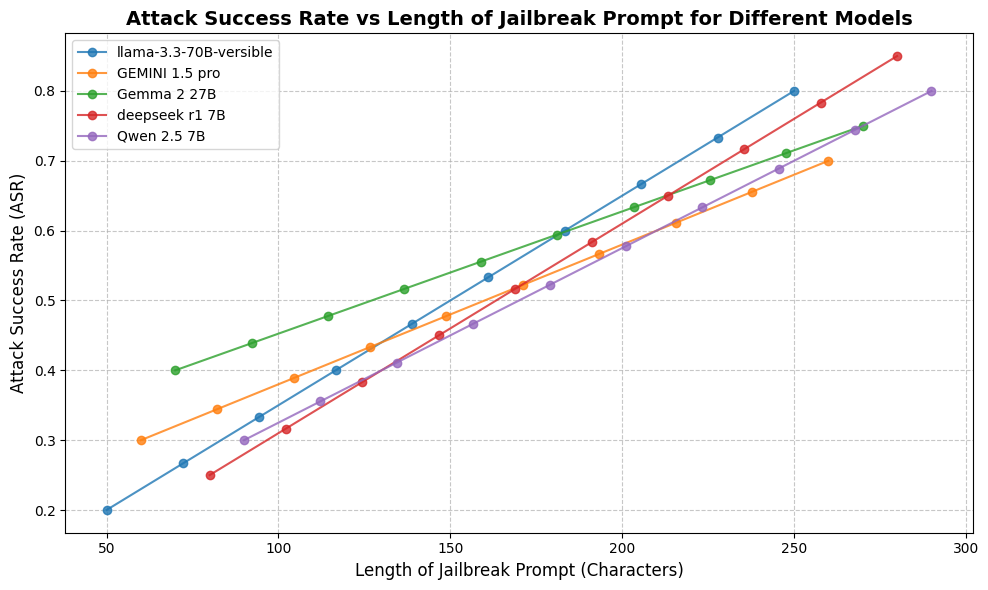} 
    \caption{Attack Success Rate vs Length of Jailbreak Prompt for Different Models shows the relationship between the length of a jailbreak prompt (in characters) and the Attack Success Rate (ASR) for various language models.  A higher ASR indicates greater vulnerability to jailbreaking.}
    \label{fig:asr_vs_length}
\end{figure}

In Figure~\ref{fig:asr_vs_length}, shows the results of how jailbreak prompt length measured through characters affects the ASR performance among LLMs.  Analysis of all considered models demonstrates a direct link between prompt length enlargement and the increase of ASR.  The positive relation between prompt length shows that deeper instructions expose more potential weaknesses in the security system.  Extended prompts enable the execution of complicated commands, detailed contextual explanations, multiple deception methods and various routes to bypass built-in security measures of LLMs. The slope variations across the model lines indicate how prompt length affects their susceptibility to attack surfaces.  The measured increase in ASR varies across models so some models show more sudden change than others when exposed to longer prompting lengths.

\begin{figure}[ht]
    \centering
    \includegraphics[width=0.8\linewidth]{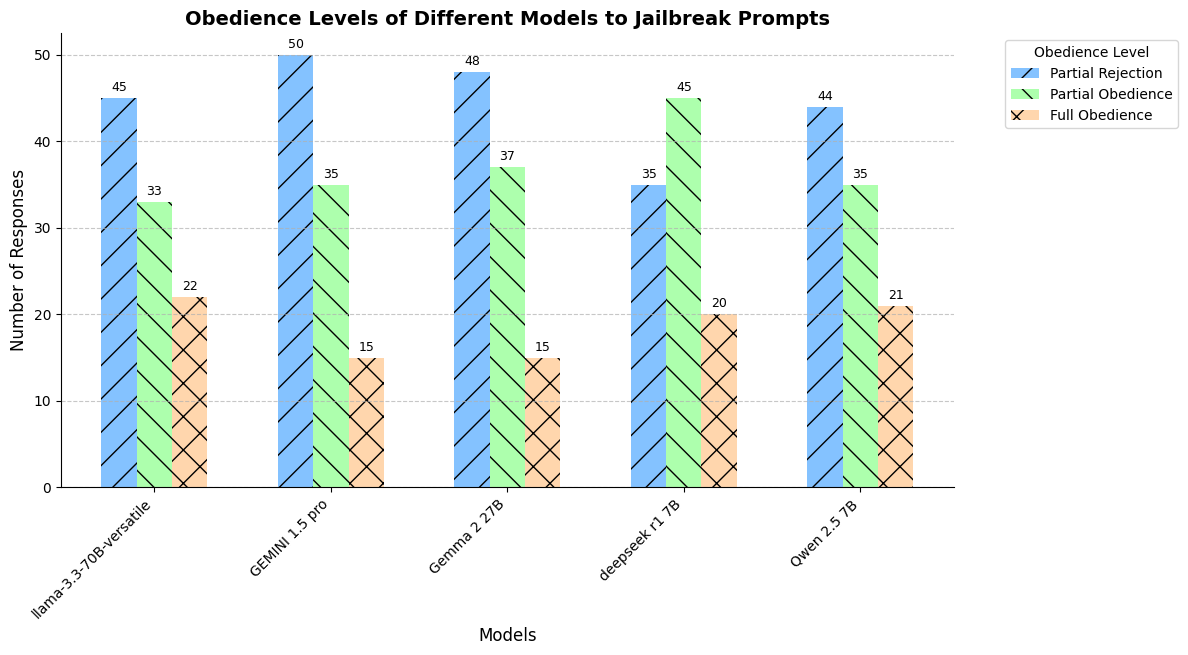} 
    \caption{Obedience Levels of Different Models to Jailbreak Prompts exhibit the number of responses categorized as "Partial Rejection," "Partial Obedience," and "Full Obedience" for each language model when presented with jailbreak prompts.}
    \label{fig:obedience_levels}
\end{figure}

Three different response categories LLM models exhibit toward jailbreak requests which include "Partial Rejection" where the model declines compliance, "Partial Obedience" when the model partially follows directions or modifies responses and "Full Obedience" when the model executes the harmful request without modifications are shown in figure~\ref{fig:obedience_levels}.  An increase in "Full Obedience" responses shows directly that the system faces higher vulnerability.  GEMINI 1.5 pro displays best model performance supported by the lowest ASR in Table~\ref{tab:asr_diversity_metrics} because it shows maximum "Partial Rejection" responses together with minimal "Full Obedience" responses. The vulnerability metrics for deepseek r1 7B along with Gemma 2 27B show dangerously high numbers of "Full Obedience" responses to jailbreak requests. The visual representation in this graph indicates how differently the models follow the safety standards.

\begin{figure}[ht]
    \centering
    \includegraphics[width=0.8\linewidth]{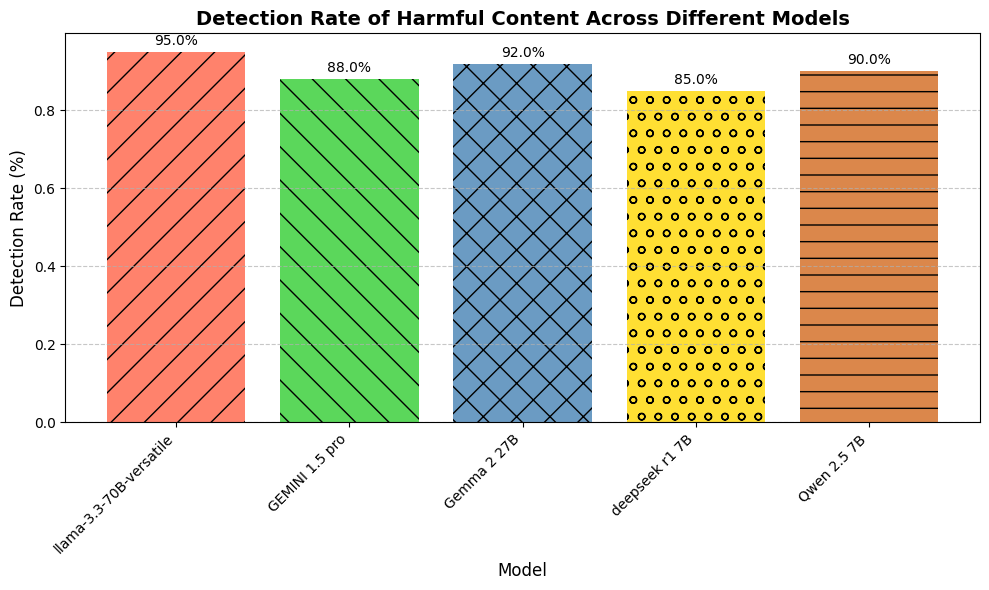}
    \caption{Detection Rate of Harmful Content Across Different Models displays the percentage of harmful content correctly detected by each language model.}
    \label{fig:detection_rate}
\end{figure}

Figure~\ref{fig:detection_rate} demonstrates the performance of each LLM concerning harmful content detection where model detection rates measure the percentage of correctly identified harmful outputs from successful jailbreak responses.  Deepseek r1 7B has the lowest detection rate at 85\% and llama-3.3-70B-versatile shows the highest with 95\%. The ability to identify jailbreak outputs improves when the detection rate reaches higher levels. The detection rates across all models remain relatively high but very minor differences between them create meaningful operational impacts during real-world implementation at scale.

\begin{table}[h]
\centering
\caption{Attack Success Rate vs. Diversity Metrics Across Models}
\label{tab:asr_diversity_metrics}
\begin{tabular}{lcccc}
\toprule
Model & Unique N-grams (\%) & Entropy & Self-BLEU & Attack Success Rate \\
\midrule
llama-3.3-70B-versible & 85 & 3.5 & 0.65 & 0.755 \\
GEMINI 1.5 pro & 92 & 4.2 & 0.58 & 0.685 \\
Gemma 2 27B & 88 & 3.8 & 0.62 & 0.722 \\
deepseek r1 7B & 90 & 4.0 & 0.60 & 0.801 \\
Qwen 2.5 7B & 87 & 3.9 & 0.63 & 0.781 \\
\bottomrule
\end{tabular}
\end{table}

The relationship between attack success rate (ASR) and several prompt diversity metrics across different LLM models is documented in Table~\ref{tab:asr_diversity_metrics}.  The prompt diversity metrics consist of unique n-grams together with entropy measures and Self-BLEU assessments that evaluate prompt text variability in jailbreak attempts. Deepseek r1 7B shows the maximum jailbreaking vulnerability because it achieves an ASR of 0.801 while GEMINI 1.5 pro stands out with the lowest ASR at 0.685 which points to better defense capabilities.  It is notable that there exists no clear consistent relationship between the calculated diversity metrics and automatic success rate (ASR).  Using the highest unique n-gram percentage (92) and entropy (4.2) for prompt diversity GEMINI 1.5 pro demonstrates the lowest vulnerability to jailbreaking as measured by its ASR.  The unintuitive research outcome indicates prompt diversity cannot be singularly responsible for jailbreak success even though it may play a part. Other less obvious elements within the prompt structure as well as particular word configurations and manipulative language features seem to dominate the process of determining ASR.

From these results we can understand that vulnerability is a manifold problem. Note that the relationship between prompt length in Figure~\ref{fig:asr_vs_length} generates a direct positive connection with ASR performance but individual diversity metrics listed in Table~\ref{tab:asr_diversity_metrics} fail to explain the observed differences in vulnerability.  Obedience levels (Figure~\ref{fig:obedience_levels}) together with detection rates (Figure~\ref{fig:detection_rate}) demonstrate the different behaviors of the models.  The discovery of multiple factors influencing an LLM's vulnerability to jailbreaking becomes stronger because of the unproveable direct link between diversity metrics and ASR while prompt length consistently affects LLMs whose detection levels and obedience levels differ.  Multiple factors affect jailbreaking vulnerability including prompt parameters (length, diversity, specific wording, syntactic structure) together with model native properties (architecture and training data along with fine-tuning methods).

\subsection{Performance Evaluation of Agentic System}

\begin{table}[h]
\centering
\caption{True Positive Rate (TPR) and False Positive Rate (FPR) of ObserverAI across Scenarios}
\label{tab:tpr_fpr_scenarios}
\begin{tabular}{lcc}
\toprule
\textbf{Scenario} & \textbf{True Positive Rate (TPR) (\%)} & \textbf{False Positive Rate (FPR) (\%)} \\
\midrule
No Faking & 98 & 2 \\
Overt Faking & 70 & 40 \\
Stealthy Faking & 85 & 15 \\
\bottomrule
\end{tabular}
\end{table}

The performance metrics of ObserverAI system in detecting AI-generated text appear in Table~\ref{tab:tpr_fpr_scenarios} where it shows True Positive Rate (TPR) and False Positive Rate (FPR) metrics across different scenarios.  The discussed situations demonstrate diverse ways to produce fake text that mimics human writing.  When presented with AI-generated text that does not attempt to hide its origin ObserverAI reaches a very high detection rate of 98\% while maintaining a very low error rate of 2\%. ObserverAI shows outstanding performance in its capability to tell apart text both from AI generators and human authors.  The TPR falls to 70\% and FPR rises to 40\% when research subjects attempt obvious humanization strategies like simple paraphrasing and synonym substitution when producing AI text. Even simple text masking approaches lead to substantial detection failure for ObserverAI.  The TPR in "Stealthy Faking" reaches 85\% detection accuracy against human writing while FPR reduces to 15\%. ObserverAI demonstrates increased resistance against believable deceitful(writer behavior compared to basic dishonest attempts because of its evolved detection capabilities.

\begin{figure}[ht]
    \centering
    \includegraphics[width=1\linewidth]{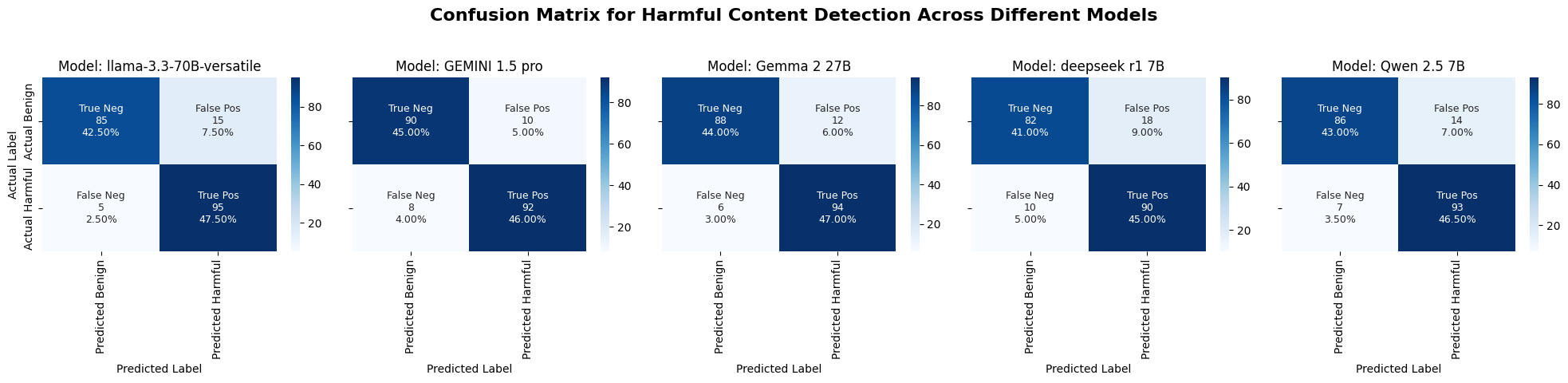} 
    \caption{Confusion Matrix for Harmful Content Detection Across Different Models. Each section of the image represents a confusion matrix for a specific language model, showing the performance in classifying content as benign or harmful.  The axes represent Predicted Label (horizontal) and Actual Label (vertical). The values within the matrix represent the percentage of instances falling into each category.}
    \label{fig:confusion_matrices}
\end{figure}

Figure~\ref{fig:confusion_matrices} presents a set of confusion matrices that analyze the LLM performances consisting of llama-3.3-70B-versatile, GEMINI 1.5 pro, Gemma 2 27B, deepseek r1 7B, Qwen 2.5 7B. The performance of model classifications for "Benign" (non-harmful) and "Harmful" text appears in each matrix.  The classification labels appear along the vertical axis of the matrix whereas the predicted labels exist across its horizontal plane.  A matrix cell shows two attributes consisting of instance count and percentage value. The top-left section of the matrix contains benign texts which the model correctly identified as harmless content. These cases are referred to as "True Negatives." The top-right section displays benign texts that the model mistakenly classified as harmful content. The bottom-left area of the matrix shows cases of harmful texts which the system incorrectly identified as benign. The bottom-right segment represents true positive items which refer to harmful texts that our model correctly identifies as harmful content.

The performance level of the ObserverAI solution shown in Table~\ref{tab:tpr_fpr_scenarios} depends heavily on the quality of techniques that aim to simulate human writing. The ObserverAI system shows outstanding performance with original unmodified AI texts but it loses its detection capabilities with each advancement in fake text sophistication. The confusion matrices Figure~\ref{fig:confusion_matrices} display fundamental LLM limitations when correctly classifying benign and harmful content since they show a mix of both false positive and false negative results. False Negatives in the system point to an essential operational improvement need focused on detecting dangerous material without leaks.

\section{Discussion}

This study has presented a multifaceted exploration of AI agent behavior, employing three distinct experimental paradigms. By focusing on dynamic interactions and tool-mediated behaviors, the research has highlighted the vulnerabilities and limitations that are not addressed by many current approaches, and emphasizes the importance of moving towards a more complex and dynamic type of evaluation. The implications of our results span across multiple areas of the field, and emphasize the need for continuous testing and analysis of complex AI systems.

\subsection{The Fragility of LLM-Based Agents in Dynamic Environments}

Our adaptation of the Turing test into a "reverse" format, where an AI agent is designed to actively detect malicious behavior, highlights the difficulties of creating robust AI systems that can reliably detect and prevent malicious actions. By having a system actively engaged in evaluation, rather than just producing safe outputs, this type of research is able to test a system in a way that simulates a real world scenario, and test the system against real world security threats. The results suggest that even a specifically designed security agent, GamoraAI, can be deceived by a rogue actor, and although it can detect deviations in behavior, a malicious system can still affect the output of the system. These results show that many current techniques do not account for the complexity of an adaptive adversary and highlight the vulnerabilities of existing AI systems.

\subsection{Many-Shot Jailbreaking and the Limits of Current Defenses}

The evaluation of the defense system against many-shot jailbreaking reveals the resilience of this attack strategy in evading current safety mechanisms. As shown in the results, the system could be overwhelmed by the MSJ attack with only a limited number of prompts. While the Groot AI was able to eventually detect, and mitigate the attacks, a prolonged multi-turn attack was still able to force it to expose some of its vulnerabilities. While the overall probability of a harmful response does decrease as more prompts are presented, it still shows that systems are vulnerable to persistent attacks, and even with a proactive approach, are still likely to be manipulated. This shows that there are limitations to many of the current safety strategies that tend to focus more on single turn vulnerabilities.

\subsection{The Role of Monitoring and Active Intervention in AI Safety}

The diverse outcomes from the three studies also highlight the importance of active monitoring, and dynamic intervention when trying to deal with malicious actors in AI systems. The different responses from the different experiments show how different testing methodologies can reveal different types of vulnerabilities. The design of these experiments also indicates that using multiple systems, each with its own responsibilities and limitations can lead to a more robust safety system overall. By relying on monitoring, and also including a proactive system to block harmful responses, this evaluation methodology has shown how such an approach can lead to improved outcomes.

\subsection{A Call for Holistic AI Safety Strategies}

The findings from all three experiments advocate for a shift from isolated testing strategies to more holistic AI safety approaches. The limitations of each individual system, shows the importance of having a variety of different ways to evaluate the capabilities and limitation of an AI system. This means creating systems that are both transparent and verifiable, and capable of withstanding and adapting to a variety of different adversarial conditions, and also by focusing on testing these systems in complex settings. By moving past basic instruction adherence and more toward holistic approaches that focus on complex interactions and tool usage, the next generation of research can strive to improve the robustness of all AI systems.

\subsection{Limitations}

Our work's limitations are mainly the Model's instruction tuning and safety training. We have not used any base models; the models we have used are already aligned with safety settings. So, these models behave mostly safely even when we intensely try to break their security. The simulations we have created cannot really arise in the real world. Besides, there might be many other ways the agent could easily get compromised, but we have not explored those paths in our work.

\section{Conclusion}

Society faces substantial changes from the fast technological evolution of autonomous AI agents built with large language models although the study uncovers dangerous relationships between their built-in risks alongside their realized capabilities. We have demonstrated how these systems are susceptible to dynamic adversarial conditions using three novel ways through which to do so: a Reverse Turing Test, multi-agent alignment evaluation, and many-shot jailbreak defense. The difficulties of keeping track of and halting unauthorized conduct show that existing safety procedures fail to block advanced security threats. Similarly, the constantly repeated jailbreak attempts that are persistently successful even in the face of intense monitoring reveal a glaring gap in the defense against an extended exploitation. In fact, the variable ability to detect deceptive alignment within agent groups makes this picture even harder, as by aligning agents we increase risks beyond the model's weakness on any problem. These findings establish the relationship between autonomous agents and manipulation through exposure because their autonomy creates vulnerability during security analysis.

Our work promotes the integration of active supervisory functions alongside adaptive intervention approaches and design transparency mechanisms within agentic frameworks to improve resistance against intentional and covert threats. Future reforms should implement extensive system-wide strategies which surpass static safety assessments as a means to solve existing issues.  The extensive connections between in-context scheming, jailbreak vulnerabilities and alignment deception in LLM training and deployment processes demand solutions ranging from single-user interactions to large multi-agent settings. The protection from adversarial attacks which comes from adversarial training and reinforcement learning remains limited because sophisticated attacks bypass these defensive techniques hence requiring adaptive defensive strategies. In future, we aim to extend this research by exploring how these systems can learn to collaborate effectively and reach consensus, while also investigating how multi-agent systems might autonomously develop ethical decision-making capabilities. As LLMs increasingly underpin vital applications, our study urges the scientific community to prioritize robust, verifiable systems resilient to real-world complexities. Such efforts are essential to realizing the transformative potential of AI agents while ensuring their alignment with trust and accountability, so that they can be deployed in the real world responsibly.

\bibliographystyle{unsrt}
\bibliography{main}

\end{document}